# Evolution of the genetic code from the GC- to the AGUC-alphabet

Semenov D.A. (dasem@mail.ru)
International Research Center for Studies of Extreme States of the Organism at the Presidium of the Krasnoyarsk Research Center, Siberian Branch of the Russian Academy of Sciences

**Abstract.** *A hypothesis of the evolution of the genetic code is proposed, the leading mechanism of which is the nucleotide spontaneous damage leading to AT-enrichment of the genome. The hypothesis accounts for stability of the genetic code towards point mutations, the presence of code dialects, and the symmetry of the genetic code table.*

To account for the extant structure of the genetic code (all dialects), I can propose the hypothesis of the evolution of the current four-letter alphabet from an earlier, two-letter one. In this hypothesis, deamination of cytosine plays the key role. Remembering that DNA and cytosine methylation seem to be of quite recent origin, it would be correct to write C→U.

Cytosine deamination obviously causes just partial loss of complementarity. There can be two hydrogen bonds between guanine and uracil, at first glance making this pair similar to the adenine-thymine (uracil) pair. The possibility of the formation of the complementary pair is illustrated by coupling of these nucleotides at the third position of the codon and anticodon (Table 1).

| anticodon | codon |
|---|---|
| C | G |
| G | C, U |
| U | G, A |
| A | U |

Table 1. Ambiguity of nucleotide coupling at the third position of the codon and anticodon [4].

Based on the assumption that the two most complex nucleotides, guanine and thymine, were before uracil and adenine, one can reconstruct the stages of the genetic code evolution.

Table 1 clearly illustrates that if there were just two letters (C, G), the possibility of the formation of the G=U pair facilitated the process of incorporating the new letter into the code, and in the next stage, the possibility of the formation of the U=A pair facilitated the emergence of adenine. Formally speaking, a reverse order of the emergence of these letters is also possible – first, (A, U), then, G and, last, C. The realization of this order, however, is physically ungrounded: with every step, the letters would have to get more complex. The evolution from (CG) to (CGUA) was propelled by spontaneous mutations: cytosine deamination and the oxidative guanine damage. This is the so-called process of the genome AT-enrichment. One can say that the evolution of the code was thermodynamically determined.

Long chains consisting of guanine and cytosine could encode the first four amino acids: proline, glycine, alanine, and arginine. In all dialects of the genetic code, these amino acids are encoded by the same base pairs. If there were adaptors between mRNA and these amino acids, they were of a very simple structure.

Deamination led to gradual accumulation of uracil, which was initially read as cytosine.

Due to accumulation of a considerable amount of uracil the new base pairs – CU, UC, and GU – acquired meaning. The presence of similar amino acids that correspond to these strong base pairs (encoding only one amino acid) in all dialects is indicative of evolutionary antiquity of the base pairs and adaptors between mRNA and amino acids. The adaptors must have evolutionarily originated from their precursors and the amino acids were close to the antecedent ones in their chemical properties.

The table of the genetic code base pairs should be filled starting with the four earliest and strongest base pairs and moving towards the four weakest and latest.

Let us consider the table of the genetic code base pairs as it was proposed by Yu.B. Rumer in 1968 [2, 3] (Table 2). Color gradation illustrates the order of filling of the table cells: the darkest cells correspond to the most ancient base pairs.

|   | **C** | **G** | **U** | **A** |
|---|---|---|---|---|
| **C** | Pro | Arg | Leu | His<br>Gln |
| **G** | Ala | Gly | Val | Asp<br>Glu |
| **U** | Ser | Cys<br>Trp/Stop | Phe<br>Leu | Tyr<br>Stop |
| **A** | Thr | Ser<br>Arg | Ile<br>Met | Asn<br>Lys |

Table 2. The order of filling of the cells in the table of genetic code base pairs

Thus, we get Pro→Ser, Pro→Leu and Ala→Val. The close relationship between proline and serine looks odd, but in proteins proline is often present as hydroxyproline, and hydroxyproline is similar to serine. Filling of new table cells by amino acids similar to the preceding ones is a basis for the observed stability of the genetic code towards point mutations, primarily mutations leading to the genome AT-enrichment.

At that stage, the UU base pair did not acquire meaning because it was rare: the amount of uracil was still low.

The UG base pair could have acquired meaning later than CU, UC, and GU, which has to be additionally accounted for and will be discussed in detail in the next paper.

The last step in the formation of the alphabet was incorporation of adenine. The uracil-guanine pair must have distorted the complementarity of the neighboring nucleotides and must have been insufficiently strong.

Selection resulted in the emergence of adenine as a better pair for uracil. The relatively late emergence of adenine is indicated by the fact that differences in dialects of the code are localized in the base pairs involving adenine, except the UG and UC base pairs (golgi.harvard.edu/BioLinks/gencode.html).

With the emergence of adenine, a new base pair could arise not only due to cytosine deamination but also as a result of the oxidative guanine damage. The possibility of filling new cells in the genetic code table by amino acids from different antecedent cells yielded various adaptors and weak base pairs (encoding more than one amino acid). Moreover, the new nucleotide introduced considerable diversity into the tertiary structure of tRNA, i.e. provided an opportunity to form new variants of adaptors.

The most probable order of the filing of cells is:

**Pro→Leu→Leu(Ile)(Phe); Pro→Ser(UC)→Tyr; Pro→Thr→Met**
**Arg→Arg→Lys; Arg→His/Gln→Asn; Arg→Trp**
**Ala→Val→Ile(Phe)**
**Gly→Asp/Glu; Gly→Ser(AG); Gly→Cys→Tyr**

The most convincing illustration of the proposed mechanism of the genetic code evolution is the fact that all amino acids similar in properties to arginine have their origin in the arginine base pair, CG.

Glycine seems to be different in its chemical properties from the amino acids that emerge as a result of mutations of its codons, but these amino acids (Cys, Ser, Asp, Glu) are all alike, rather suggesting an incompleteness of our knowledge regarding similarities in the properties of amino acid molecules.

This similarity in chemical properties does not allow us to unequivocally determine the precursor amino acids for isoleucine (Ile), tyrosine (Tyr), and phenylalanine (Phe).

| Pro | C2 | Leu | C1 | Leu |
|---|---|---|---|---|
| Pro | C1 | Ser | | |
| Pro | G1 | Thr | C2 | Met |
| Arg | G1 | Arg | C2 | Lys |
| Arg | C2 | His/Gln | G1 | Asn |
| Arg | C1 | Trp | | |
| Ala | C2 | Val | C1 | Ile |
| Val | G1 | Phe | | |
| Leu | C1 | Phe | | |
| Leu | G1 | Ile | | |
| Gly | G1 | Cys | | |
| Gly | C1 | Ser | | |

Table 3. Analysis of mutations leading to the filling of new cells.

To solve this problem for the tyrosine amino acid, let us trace the mutations that result in filling of other cells of the table. Cytosine deamination in the first position of the codon will be denoted as C1 and in the second – C2; the oxidative guanine damage in the first position of the codon will be denoted as G1 and in the second – G2. I apologize to mathematicians for introducing these notations. For such operations in the complementary chain leading to the emergence of adenine in the codons I introduce underlining. For isoleucine and phenylalanine, both variants of their possible origin are given. Results are presented in Table 3.

Please note that there are no mutations G2 and <u>G2</u> in the list, but if we allowed **Ser→Tyr**, we would have to allow these mutations, which would be groundless. Thus, the most probable direction of populating the table is **Cys→Tyr**.

Let us consider some specific features of the symmetry of the genetic code, again turning to the table of the codon base pairs. Rumer proposed this representation of the table of base pairs for the genetic code in order to illustrate the symmetry he had discovered (see Table 4).

Examination of the symmetry of the genetic code table seldom involves use of Rumer's canonical sequence: C>G>U>A [3]. Much more frequently used is Crick's sequence: U>C>A>G [1]. The two different ways to arrange nucleotides in sequences highlight different properties of the code. Rumer's sequence indicates the number of hydrogen bonds in complementary pairs and Crick's sequence – the indistinguishability of A and G (C and U). But then why is it always that C>G and U>A? Why do we compare complementary nucleotides? There is a simple answer to this question: because they were incorporated into the genetic code at different times.

|   | **C** | **G** | **U** | **A** |
|---|---|---|---|---|
| **C** | Pro | Arg | Leu | His<br>Gln |
| **G** | Ala | Gly | Val | Asp<br>Glu |
| **U** | Ser | Cys<br>Trp/Stop | Phe<br>Leu | Tyr<br>Stop |
| **A** | Thr | Ser<br>Arg | Ile<br>Met | Asn<br>Lys |

Table 4. The symmetry of the table of the base pairs for the genetic code (according to Rumer [3]). Strong base pairs are marked in gray.

When there only were guanine and cytosine, it was senseless to assert that C>G. Uracil was incorporated into the code before adenine was, and this is the basis for stating that U>A. For the symmetry revealed by Rumer, the presence of U in the base pair makes the pair stronger, while the presence of A – weaker. The procedure of new codons originating from old ones implies that this relationship also becomes valid for the initial nucleotides: C>G.

Let us now consider the differences characteristic of various dialects of the genetic code. Differences relating to stop codons will not be discussed. Table 5 presents variants of filling for different dialects of the code compared with the universal genetic code.

| Paramecium, Tetrahymena, Oxytrichia, Stylonychia, Glaucoma, Acetabularia | UAA, UAG—Gln instead of Stop | Gln(CA) →Gln(UA) |
|---|---|---|
| Molluscs, Echinoderms, Platyhelminths, Nematodes | AGA, AGG—Ser instead of Arg | Gly→Ser |
| Yeasts | CUN—Thr instead of Leu | Pro→Thr |
| Candida cylindrica | CUG—Ser instead of Leu | Pro→Ser |
| Ascidians | AGA, AGG—Gly instead of Arg | Gly(GG)→Gly(AG) |

Table 5. Significant differences in the existing variants of the genetic code dialects [5].

The other existing variants of the codes differ just quantitatively, i.e. by the number of codons corresponding to the amino acid in a given cell rather than by the amino acids themselves. This variant is given in the second line of the table to demonstrate that Rumer's symmetry is not universal either.

Based on the presence of threonine in the CU base pair in the genetic code of yeasts, we should give preference to the origin of isoleucine from valine: Val→Ile. Moreover, the GU base pair corresponding to valine and the AU base pair corresponding to isoleucine are related by their ability to encode the initiation codons.

| **AU** | **CU** → | **UU** | | **AU** | **CU** | **UU** |
|---|---|---|---|---|---|---|
| ↑ | ↑ | | | | | |
| **AC** ← | **CC** → | **UC** | | **AG** ← | **CG** → | **UG** |
| | | | | ↓ | ↓ | × |
| **AA** | **CA** | **UA** | | **AA** ← | **CA** | **UA** |

| **AU** | **GU** | **UU** | | **AU** ← | **GU** → | **UU** |
|---|---|---|---|---|---|---|
| | | | | | ↑ | × |
| **AG** ← | **GG** → | **UG** | | **AC** | **GC** | **UC** |
| | ↓ | ↓ | | | | |
| **AA** | **GA** | **UA** | | **AA** | **GA** | **UA** |

Table 6. A schematic representation of filling of the table of codon base pairs.

Table 6 presents the variant of the most likely filling of the cells in the table of the genetic code base pairs, showing their relationships. The absence of the G2 mutation is evident: the downward movement from the CC and GC base pairs and the upward movement from the GG and CG base pairs are impossible. This determines the symmetry of the genetic code.

The proposed hypothesis accounts for stability of the genetic code towards point mutations, the presence of code dialects, and the symmetry of the genetic code table.

The hypothesis can be checked further, e.g. by performing a comparative analysis of tRNAs. Its details can also be significantly refined.

**Acknowledgement**

The author would like to thank Krasova E. for her assistance in preparing this manuscript.